\documentclass[aps,pre,superscriptaddress,twocolumn,showpacs,nofootinbib,amsmath]{revtex4}
\usepackage{graphicx}

\newcommand{\be}{\begin{equation}}
\newcommand{\ee}{\end{equation}}
\newcommand{\bse}{\begin{subequations}}
\newcommand{\ese}{\end{subequations}}
\newcommand{\bea}{\begin{eqnarray}}
\newcommand{\eea}{\end{eqnarray}}

\newcommand{\fra}[2]{\hbox{${#1\over #2}$}}
\newcommand{\comment}[1]{}

\begin{document}

\title{
Modified Kuramoto-Sivashinsky equation:
stability of stationary solutions and the consequent dynamics}

\author{Paolo Politi}
\email{paolo.politi@isc.cnr.it}
\affiliation{Istituto dei Sistemi Complessi,
Consiglio Nazionale delle Ricerche, Via Madonna del Piano 10,
50019 Sesto Fiorentino, Italy}

\author{Chaouqi Misbah}
\email{chaouqi.misbah@ujf-grenoble.fr} \affiliation{Laboratoire de
Spectrom\'etrie Physique, CNRS, Universit\'e J. Fourier, Grenoble 1, BP87,
F-38402 Saint Martin d'H\`eres, France}

\date{\today}

\begin{abstract}
We study the effect of a higher-order nonlinearity in the standard
Kuramoto-Sivashinsky equation: $\partial_x \tilde G(H_x)$. We find
that the stability of steady states depends on $dv/dq$, the
derivative of the interface velocity on the wavevector $q$ of the
steady state. If the standard nonlinearity vanishes, coarsening is
possible, in principle, only if $\tilde G$ is an odd function of
$H_x$. In this case, the equation falls in the category of
the generalized Cahn-Hilliard equation, whose dynamical behavior
was recently studied by the same authors. 
Instead, if $\tilde G$ is an even function of $H_x$, we show
that steady-state solutions are not permissible.
\end{abstract}



\maketitle
\section{Introduction}
One of the most prominent and generic equations that arises in
nonequilibrium systems is the Kuramoto-Sivashinsky
(KS)\cite{Nepo,KS,Sivashinsky} equation:
\be
H_t + H_{xxxx} + H_{xx} + H H_x =0 ,
\label{eq_KS}
\ee
where $H$ is some scalar function (like the slope of a
one-dimensional growing front), and differentiations are
subscripted. The linear stability analysis of the KS equation (by
looking for solutions in the form of $e^{iqx+\omega t}$) yields
$\omega=q^2-q^4$.
The (linearly) fastest growing mode has a
wavenumber given by $q_u=1/\sqrt{2}$ (obtained from $\partial _q
\omega=0$). For a large box size the KS equation is known to exhibit
spatiotemporal chaos. The chaotic pattern statistically selects a
length scale which is close to $2\pi/q_u$: in fact, the structure factor
$\langle |H_q|^2\rangle $, where $\langle\cdots\rangle$
designates the average over many runs and $H_q$ is the Fourier
transform of $H$, exhibits a maximum around $q=q_u$.
Other nonequlibrium equations are known, however, to
exhibit different dynamical behaviors: just to limit to
one-dimensional systems, we may have coarsening, a diverging
amplitude with a fixed wavelength, a frozen pattern, travelling waves,
and so on.

An important issue is the recognition of
general criteria  that enable to predict whether or not
coarsening takes place within a class of nonlinear equations,
without having to resort to a forward time dependent calculation.
In recent works~\cite{Politi1,Politi2,CKS} we have considered
several classes of one-dimensional Partial Differential Equations
(PDE), having the form $H_t = {\cal N}[H]$, where ${\cal N}$
is a nonlinear operator acting on the spatial variable $x$.

\begin{figure}
\includegraphics[width=7cm,clip]{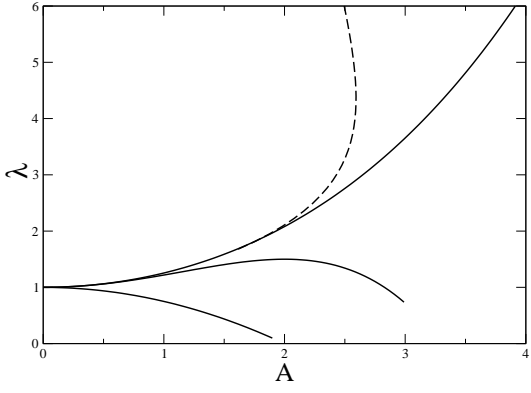}
\caption{The periodic stationary solutions of some classes of nonlinear
equations satisfy a Newton-type equation, $\tilde H_{xx} + V(\tilde H)=0$,
and the resulting $\lambda(A)$ is a single-value function.
According to the explicit form of $V(\tilde H)$, we may have the
different curves shown as full lines. The dashed line, which
displays a fold, corresponds to a non single-value function.
It comes out in the Kuramoto-Sivashinsky and in the
Swift-Hohenberg equations. Units on both axes are arbitrary.
}
\label{lambda_A}
\end{figure}

Sometimes, even in the presence of strong nonlinearities, the search
for steady states reduces to solving a Newton-type equation, $\tilde
H_{xx} + V(\tilde H)=0$, where $\tilde H$ is some function of $H$.
In these cases, $\lambda(A)$, giving the dependence of the
wavelength $\lambda$ of the steady state on its amplitude $A$, is a
one-value function (see Fig.~\ref{lambda_A}, full lines), and the
criterion for the existence of coarsening is expressed in terms of
the derivative $\lambda'(A)$.

It has been shown that
$\lambda '(A)$ has minus the sign of the phase diffusion
equation. Thus $\lambda '(A)>0$ corresponds to a branch which is
unstable with respect to the phase of the pattern,
entailing thus coarsening.
The situation is more complicated when $\lambda(A)$ exhibits
a fold (see Fig.~\ref{lambda_A}, dashed line). This event occurs, e.g.,
in the KS equation and in the Swift-Hohenberg equation.
As for the KS equation,
which is the topic of this paper, Nepomnyashchii~\cite{Nepo}
has shown that the forearm part of the curve $\lambda(A)$
with positive slope, $\lambda'(A)>0$, is an unstable branch.
This result holds only for the pure KS equation, however.

The aim of this paper is the following. (i) Firstly we shall extend
the result of Nepomnyashchii~\cite{Nepo} to a generalized form of
the KS equation, which includes higher order nonlinearities. As a
way of example, the next leading term in the KS equation
($H_xH_{xx}$) has been analyzed in~\cite{Ihle}, and it has been
shown that this term significantly affect dynamics; for example the
profile may exhibit deep grooves. We shall  consider a more general
form of the modified KS equation by adding a term like $\partial _x
\tilde G (H_x)$ (this includes  as a particular case the term
$H_xH_{xx}$). We find that the stability of the steady state
solutions depends on $v'(q)$, where $v$ is the average interface
velocity. (ii) If the standard KS nonlinearity vanishes (no $HH_x$
term), then there is coarsening if $\tilde G$ is an odd function. In
this case a mapping of the equation onto a generalized Cahn-Hilliard
equation is straightforward. (iii) If $\tilde G$ is even (still in
the absence of the standard nonlinearity), we show that there exists
no steady-state periodic solution, as attested by numerical
simulations for $\tilde G= H_x^2$~\cite{CMV}.

\section{The modified KS equation}

\subsection{The method}

We study the following equation:
\be
H_t +c_4 H_{xxxx} +c_2 H_{xx} + \alpha H H_x +\beta \partial_x \tilde G(H_x)=0 ,
\ee
which
reduces to the standard Kuramoto-Sivashinsky equation when $\tilde
G=0$. A rescaling of $x,t$ and $H$ always allows one to reduce the
equation to a one parameter equation, which can be absorbed
into a redifinition of $\tilde G$. However, for the sake of
clarity, we do not get rid of $\alpha$, so that we write
\be
H_t + H_{xxxx} + H_{xx} + \alpha H H_x + \partial_x \tilde G(H_x)=0 .
\label{eq_KSm}
\ee

It is also useful to rewrite
Eq.~(\ref{eq_KSm}) using the variable $u$, with $u_x=H$:
\be
u_t + u_{xxxx} + u_{xx} + \fra{\alpha}{2} u_x^2 + \tilde G(u_{xx})=0 ,
\ee
where the integration constant $v_0$ can be canceled out by the
transformation $u(x,t)\to u(x,t) -v_0 t$.

Within the $H-$formulation, the average velocity $d\langle H\rangle/dt$
vanishes, because Eq. (\ref{eq_KSm}) has the conserved form
$H_t=- \partial_x (\dots)$. Within the $u-$formulation,
\be
{d \langle u\rangle\over dt} = -
\left[{\alpha\over 2} \langle u_x^2 \rangle +\langle \tilde G (u_{xx}) \rangle
\right]
= v .
\ee

We start from a stationary solution of period $q$, $H(x)$, and perturb it with
adding $h(x)\exp(-\omega t)$. The function $h(x)$ therefore satisfies
the linear equation
\be
h_{xxxx} + h_{xx} + \alpha (hH)_x + (h_x G(H_x))_x = \omega h ,
\label{eq_hx}
\ee
where $G=\tilde G'$ and
whose coefficients are periodic with period $\lambda=2\pi/q$.
According to the Floquet-Bloch theorem, the solution has the form
$h(x)=\exp(iKx)F(x)$, where $F(x)$ has the same period as $H(x)$.

We are interested in weak modulations of long wavelength, i.e.
with $K\ll q$. It is therefore convenient to introduce the reduced
wavevector $Q=K/q$ and the phase $\phi=qx$.
The equation determining the steady state $H(\phi)$ is
\be
q^4 H_{\phi\phi\phi\phi} + q^2 H_{\phi\phi} +\alpha q HH_\phi +
q^2 H_{\phi\phi} G(qH_\phi) =0
\label{eq_stat}
\ee
and Eq.~(\ref{eq_hx}) reads
\bea
q^4 h_{\phi\phi\phi\phi} +q^2 h_{\phi\phi} + q\alpha (hH)_\phi +
q^2 (h_\phi G(qH_\phi))_\phi \nonumber \\
\equiv {\cal L}h = \omega h ,
\label{eq_hphi}
\eea
with $h=\exp(iQ\phi)F(\phi)$.

The final step is to expand both $\omega$ and $F(\phi)$ in powers
of $Q$,
\bea
F(\phi) &=& F_0(\phi) + Q F_1(\phi) + Q^2 F_2(\phi) + \dots \\
\omega &=& \omega_0 + Q \omega_1 + Q^2 \omega_2 + \dots
\eea
and to solve Eq.~(\ref{eq_hphi}) at the lowest orders in $Q$.

\subsection{Zero order}

The differential equation determining the steady state $H(x)$ does not
depend explicitely on $x$, so that $H(x+x_0)$ is a solution as well.
This symmetry implies that ${\cal L}h=\omega h$ (see Eq.~(\ref{eq_hphi}))
is solved by $h=H_\phi$ and $\omega=0$,
as can be easily checked by taking the $\phi$ derivative of
Eq.~(\ref{eq_stat}). Since the zero order equation is simply
\be
{\cal L}F_0=0 ,
\ee
we obtain $F_0=H_\phi$.

\subsection{First order}

The equation for $F_1$ reads
\bea
&&{\cal L}F_1 =
\label{eq_F1} \\
&&\omega_1 H' -iq[ 4q^3 H'''' +2q(1+G)H''+ (\alpha H +q^2H''G')H' ] , \nonumber
\eea
where we have used the shorthands
$H'=H_\phi,H''=H_{\phi\phi},\dots$~. If we differentiate
Eq.~(\ref{eq_stat}) with respect to $q$, we get a similar equation,
\be
{\cal L}H_q = - (4q^3 H'''' +2qH''+\alpha HH'+3\beta q^2H'H'') .
\label{eq_Hq}
\ee

The comparison of Eqs.~(\ref{eq_F1},\ref{eq_Hq}) suggests to look for
$F_1$ under the form $F_1=iqH_q+c$, where $c$ is a constant.
We easily find
$c=\omega_1/(\alpha q)$, so that
\be
F_1={\omega_1\over\alpha q} + iqH_q .
\label{F1}
\ee
This result shows that in the absence of the
standard nonlinearity, $\alpha=0$, $\omega_1$ should vanish whatever
$\tilde G$ is.

We might have started with an even more general Eq.~(\ref{eq_KSm}),
replacing the standard nonlinearity $(\alpha HH_x)=\partial_x (
\fra{\alpha}{2} H^2)$ with
$\partial_x \tilde P(H)$. The term $\alpha HH'$ in
the right hand sides of Eqs.~(\ref{eq_F1}) and (\ref{eq_Hq}) would be replaced
by $P(H)H'$, with $P=\tilde P'$. However, in the general case $P'$ is not a
constant: therefore, it is not possible to look for a solution $F_1=iqH_q + c$.

\subsection{Second order}

The equation for $F_2$ has the form
\bea
&&{\cal L}F_2 +q^2 (1+G)(-F_0+2iF_1') +(\alpha qH +q^3H''G')iF_1 \nonumber\\
&&= \omega_2 F_0 + \omega_1 F_1 ,
\label{eq_F2}
\eea
or
\bea
{\cal L}F_2 &=& \omega_2 H' + iq\omega_1 H_q+q^2 H' +2q^3 H_q'
-i\omega_1 H \nonumber \\
&& -i(\omega_1/\alpha) q^2 H'' G'(qH')\nonumber\\
&& +\omega_1^2/(\alpha q) +q^2 G(qH')H' + 2q^3 G(qH') H'_q\nonumber \\
&& +\alpha q^2HH_q  +q^4 H'' G'(qH') H_q .
\eea

Now we take the $2\pi$-average of the previous equation, getting
\bea
{\omega_1^2\over\alpha q} +q^2\langle G(qH')H'\rangle
+2q^3\langle G(q(H') H'_q\rangle\nonumber \\
+\alpha q^2 \langle HH_q\rangle +q^4 \langle H''G'(qH')H_q\rangle =0 .
\eea

Finally, we obtain
\be
\omega_1^2 = -\alpha q^3 {d\over dq} \left[{\alpha\over 2} \langle H^2\rangle
+\langle \tilde G(qH')\rangle \right] \equiv \alpha q^3 {d\over dq} v .
\ee

This result proves that $\omega_1=0$ if $\alpha=0$, whatever the
function $\tilde G$ is. Consider the case $\alpha\ne 0$ (we are at
liberty to choose $\alpha>0$).  Since $\omega<0$ signals an
instability, one sees that if ${d\over dq} v>0$, the periodic
solution is unstable, because there is a real solution $\omega_1<0$.
This generalizes the result of \cite{Nepo}, obtained for the pure KS
equation, to the higher order KS equation. It is only in the pure KS
limit that the spectrum of stability is related to the slope of the
steady amplitude. In the higher order equation, however, this ceases
to be the case. Instead we should replace the amplitude by the drift
velocity, a quantity which can be still obtained from pure steady-state
considerations.

\subsection{Determination of $\omega_2$}

The determination of $\omega_2$ implies the resolution of the
differential equation
\be
{\cal L}^\dag u=0 ,
\ee
where
\be
{\cal L}u = u_{xxxx} + u_{xx} +\alpha (Hu)_x + [G(H')u_x]_x
\ee
and
\be
{\cal L}^\dag u = u_{xxxx} + u_{xx} -\alpha Hu_x + [G(H')u_x]_x .
\ee

The equation ${\cal L}u=0$ is solved by $u=H'$, but we do not know
the solution of ${\cal L}^\dag u=0$.
${\cal L}^\dag \ne {\cal L}$ because of the $\alpha$--term.
If $\alpha=0$, such term is absent and ${\cal L}^\dag = {\cal L}$.
This case is treated in the next subsection.

\subsection{The case $\alpha =0$}

If $\alpha=0$, stationary solutions are determined by
the equation
\be
H_{xxx} + H_x + \tilde G(H_x) = C ,
\ee
where $C$ is a constant. If $h=H_x$, then
\be
h_{xx} = - h -\tilde G(h) +Cx .
\ee

The equation for $H(x)$ admits periodic solutions only if we take
$C=0$ and if $\tilde G(h)$ is an odd function, so that $h(x)$ itself
is periodic and has zero average for any initial condition (such
that the solution is bounded).

In the same limit $\alpha=0$, the full PDE~(\ref{eq_KSm}) writes
\be
H_t = -\partial_x[  H_{xxx} + H_x + \tilde G(H_x) ]
\ee
and taking the spatial derivative of both terms, we get
\be
h_t = -\partial_{xx} [h_{xx} + h + \tilde G(h) ] ,
\ee
where $h=H_x$. We have therefore got a generalized
Cahn-Hilliard equation, whose dynamical behavior is
known to show coarsening if and only if the wavelength
$\lambda$ of steady states is an increasing function of their
amplitude $A$~\cite{Politi2}.
We have reobtained the same result following the method discussed
in this Section.

\section{Final remarks}

Present and recent work~\cite{Politi1,Politi2,CKS} has the main
objective to find general criteria to understand and anticipate
the dynamics of nonlinear systems by the analysis of steady state
solutions only. For some important classes of PDE, which have the
common feature of a single-value $\lambda(A)$ function, the criterion
is based on the sign of the derivative $\lambda'(A)$. In this short
note we have considered a modified, generalized Kuramoto-Sivashisnky
equation, where the $\lambda(A)$ curve is not single-value, because it
displays a fold. We have therefore established a different criterion,
based on $dv/dq$, the derivative of the interface velocity on the
wavevector $q$ of the steady state.

If the standard KS nonlinearity is absent we can say more. If the
nonlinearity $\partial_x \tilde G(H_x)$ corresponds to an even
function $\tilde G$, the equation does not support periodic
stationary solutions and this prevents coarsening in principle: we
have a pattern of fixed wavelength and diverging amplitude. Instead,
if $\tilde G$ is an odd function, the equation falls in a previous
studied class, the generalized Cahn-Hilliard equation, which can
show different behaviors according to the form of $\lambda(A)$.
Finally, it is an important task for future investigations to see
whether or not information of the types presented here and in
\cite{Politi1,Politi2} have  analogues in higher dimensions.

\end{document}